\documentclass[superscriptaddress,iop,apj]{emulateapj}
\usepackage{latexsym}
\usepackage{amssymb}
\usepackage{amsfonts}
\usepackage{threeparttable}                                                               
                            
\usepackage{amsmath}
\usepackage{bm}
\usepackage[normalem]{ulem}
\usepackage{multirow}
\usepackage{natbib} 
\usepackage{subfigure}
\usepackage{color}
\usepackage{calc}
\usepackage{amsmath,amssymb,graphicx}
\usepackage{amssymb,amsmath}
\usepackage{tensor}
\usepackage{bm}
\usepackage{microtype}
\usepackage{booktabs}
\usepackage{times}
\usepackage[varg]{txfonts}
\usepackage[colorlinks, pdfborder={0 0 0}]{hyperref}
\usepackage{subfigure}
\usepackage{lipsum}
\usepackage{breakurl}
\usepackage{graphicx}
\usepackage{pbox}

\shorttitle{A Fast Accreting Neutron Stars Population Among Low Mass X-Ray Binaries}
\shortauthors{Patruno et al.}

\begin{document}
\title{The Spin Distribution of Fast Spinning Neutron Stars in Low Mass X-Ray Binaries: Evidence for Two Sub-Populations}
\author{A. Patruno}
\affiliation{Leiden Observatory, Leiden University, P.O. Box 9513, NL-2300 RA Leiden, The Netherlands}
\affiliation{ASTRON, Netherlands Institute for Radio Astronomy, Postbus 2, NL-7990 AA Dwingeloo, The Netherlands}
\author{B. Haskell}
\affiliation{Nicolaus Copernicus Astronomical Center, Polish Academy of Sciences, ul. Bartycka 18, 00-716 Warsaw, Poland}
\author{N. Andersson}
\affiliation{Mathematical Sciences and STAG Research Centre, University of Southampton, Southampton SO17 1BJ, UK}
\begin{abstract}

We study the current sample of rapidly rotating neutron stars in both accreting and non-accreting binaries in order to determine whether the spin distribution of accreting neutron stars in low-mass X-ray binaries can be reconciled with  current accretion torque models.  We perform a statistical analysis of the spin distributions and show that there is evidence for two sub-populations among low-mass X-ray binaries, one at relatively low spin frequency, with an average of $\approx300$ Hz and a broad
  spread, and a peaked population at higher frequency with
  average spin frequency of $\approx575$ Hz. We show that the two
  sub-populations are separated by a cut-point at a frequency of
  $\approx540$ Hz. We also show that the spin frequency of radio
  millisecond pulsars does not follow a log-normal distribution and shows no evidence for the existence of distinct sub-populations. We
discuss the  uncertainties of different accretion models  and  speculate that either the accreting neutron star cut-point marks the onset of gravitational waves as an efficient mechanism to remove angular momentum or some of the neutron stars in the fast sub-population do not evolve into radio millisecond pulsars.  

\end{abstract}



\section{Introduction}
\bigskip

The fastest spinning neutron stars are found either as old radio
millisecond pulsars (RMSPs) or in their progenitor systems, the accreting
neutron stars in low mass X-ray binaries (LMXBs).  Accretion theory
predicts that the majority of neutron stars in LMXBs should
pulsate and therefore be accreting X-ray pulsars.  However, the number
of AMXPs is only 19 among $\approx200$ non-pulsating LMXBs (see
\citealt{pat12r} for a review). Ten other LMXBs are nuclear powered
pulsars (NXPs), showing short lived "burst oscillations" during
runaway thermonuclear explosions on the neutron star surface~\citep{wat12}.
\cite{tau12} proposed that when an AMXP finally turns into a radio
millisecond pulsar, the neutron star strongly decelerates and loses
about half of its rotational energy during the detachment process of
the companion from its Roche lobe. According to this scenario, radio ms pulsars should therefore be relatively slower than accreting ms pulsars.

However, this does not explain why accreting sub-millisecond pulsars
have not been found (so far). Indeed, a consequence of the scenario
proposed by ~\cite{tau12} is that the fastest RMSPs known today (like
the 716 Hz PSR J1748-2446ad,~\citealt{hes06}) should have been
spinning at more than 1000 Hz during their accreting phase.  This is
not observed in the population of accreting neutron stars, where no
object is known to spin faster than about 619 Hz~\citep{har03}.  

An interesting
scenario predicts that gravitational waves act as a "brake" on the
neutron star spin and balance the accretion torques, although it is
still unclear whether this really is the
case~\citep{has11}. \cite{bil98} and \cite{aks98} had anticipated
that a speed limit to spinning neutron stars should be expected if
gravitational waves are emitted by these systems.  As noticed by
\cite{cha03} and \cite{cha08}, at frequencies of the order of 700 Hz
or more, the braking torque applied by gravitational waves (the
strength of which scales as the 5th power of the spin frequency for deformed rotating stars) might be
sufficiently strong to balance accretion torques and prevent a further
spin up. However, it is by no means clear that we need gravitational waves to explain these systems. \cite{pat12a} first showed that the presence of a
magnetosphere with a minimum strength of $\sim10^8$ G is in principle
sufficient to explain the lack of neutron stars spinning faster than
$\sim700$ Hz. In that work the condition for the spin equilibrium set
by the disk/magnetosphere interaction was calculated for an average
mass accretion rate during an outburst. However, \cite{bha17} and
\cite{dan16} have recently shown that transient accretion with a
varying mass accretion rate can strongly affect the spin evolution of
these systems. In particular, \cite{bha17} have shown that, when this
effect is taken into account, the existence of a magnetosphere with a
strength of $10^8$ G is no longer sufficient to explain the observed spin
frequency limit.  This suggestion is  indicative but
several open issues remain, the most important ones being which
precise mechanism generates gravitational waves, whether this mechanism is
sufficient to balance accretion torques on timescales of hundred
millions of years and to what extent it can reproduce the observed spin
evolution of accreting pulsars. Moreover, alternative hypotheses
like the possible presence of a trapped disk (which substantially modifies the
long-term accretion torque) need to be considered seriously~\citep{dan16}.

Considering the overall population, 
\cite{pap14b} performed a statistical analysis of
different subsets of neutron stars (see also~\citealt{hes08} for
seminal work on the same topic). They found that radio
millisecond pulsars (RMSPs) are on average slower than nuclear powered
pulsars, a fact that could in principle be ascribed to the loss of
angular momentum during the Roche lobe decoupling phase (although, see
\citealt{dan16} for a critique of this scenario).

In this paper we perform an extended statistical analysis on an
updated sample of millisecond pulsars -- including radio, accretion and
nuclear powered systems. We find that, even though most of the
conclusions of \cite{pap14b} still hold, there is a significant
difference in the behavior of RMSPs and accreting neutron stars, even
when we consider the small sample of AMXPs alone
(Section~\ref{sec:cri}). We discuss the spin
distribution of fast spinning accreting neutron stars by dividing the observed systems into
sub-populations and performing a statistical analysis to search
for evidence of sub-populations by statistical inference. We
justify this approach in Section~\ref{sec:dis} and show the
existence of a strong peak in the distribution and a separation
cut-point above which a significant subgroup of objects seem to cluster. 
In Section~\ref{sec:pop} we verify whether there is sufficient observational evidence for the presence of a 
common underlying population for all different sub-groups of neutron stars (accreting and non accreting).
In Section~\ref{sec:obstheory} we discuss open problems
with the current torque theories, whether they are
sufficient to explain the observed distribution and what  role may be played by gravitational waves. 
We then
discuss the properties of all accreting neutron stars which are close
to the observed spin frequency limit (Section~\ref{sec:sou}).
Finally, we consider recent evidence pointing to the absence of a magnetosphere in some of these systems, and the implications of this for the spin evolution (Section~\ref{sec:mag}).

\section{Pulsar Spin Distributions}\label{sec:cri}

\cite{pap14b} considered several different neutron star groups,
specifically the binary radio millisecond pulsars, the eclipsing radio
ms pulsars (which we refer to as ``spiders'' from now on), accreting and
nuclear powered ms pulsars and a subgroup of objects that comprised
all accreting ms pulsars plus all eclipsing radio pulsars.  The
latter group was called ``transitional millisecond pulsars'' under the
assumption that all AMXPs and eclipsing radio pulsar binaries show
transitions between a radio ms pulsar state and an LMXB state (as seen
in three such systems, see~\citealt{arc09,pap13a,bas14}).  In this
work we do not consider this last group. Instead, when we  refer to  transitional ms
pulsars (tMSPs) we consider only the three systems for
which there is actual evidence of a transition.  We also consider both isolated
ms pulsars and binary ms pulsars (which form the RMSPs sample after excluding the spiders), since the
progenitor of isolated ms pulsars must have gone through episodes of
accretion (recycling) in their past history. We do this because we
want to introduce the least amount of selection bias in our samples.
Our sample of neutron stars is summarized in Table~\ref{tab:pulsars}
(see also the catalog https://apatruno.wordpress.com/about/millisecond-pulsar-catalogue/ for a complete list of objects).  
Another difference in this work is that, in contrast to
\cite{pap14b} we do not interpret our $p$-values as the probability
for the null hypothesis since it is not possible to associate any
probability for the null in hypothesis testing. 

\begin{table}
  \caption{Radio, Accreting and Nuclear Powered Millisecond Pulsar Groups. $H_0$ is the null hypothesis that the data follows a normal distribution.}
\centering
\begin{tabular}{llllll}
\hline
\hline
Group & Nr. Objects & $\langle\nu\rangle$ & Std. Dev. & Shapiro-Wilk & Reject $H_0$\\
      &   &      [Hz]           &  [Hz]  & ($p$-value)\\ 
\hline
RMSPs & 337 & 253.6 & 131.6 & 1.2e-5 & yes\\
AMXPs & 19 & 367.8 & 153.0 & 0.086 & no \\
Spiders & 61 & 366.5 & 155.2 & 0.70 & no \\
NXPs  & 10 & 502 & 123.5 & 0.017 & yes \\
LMXBs  & 29 & 414.1 & 155.5 & 0.012 & yes \\
\hline
\end{tabular}\\\label{tab:pulsars}
\end{table}

\subsection{Do Pulsar Spins Follow a Normal Distribution?}

\begin{figure}
  \centering
  \rotatebox{0}{\includegraphics[width=0.9\columnwidth]{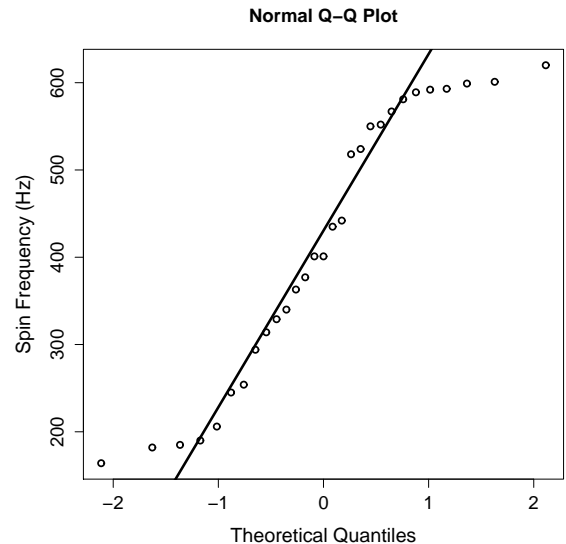}}
  \caption{Quantile-quantile plot for the LMXBs (AMXPs+NXPs) distribution. The $45^{\circ}$ oblique line is the expected theoretical trend for a normal distribution. The deviations are strongest towards the tail of the distribution.}
  \label{fig:qqplot}%
\end{figure}

Throughout this work we use the statistical package R (v 3.3.3) to
perform our analysis.  The first test we apply is a
Shapiro-Wilk (SW) normality test (package \emph{stats}) to check
whether the spin frequency of each sample is compatible with a normal
distribution (the null hypothesis $H_0$ being that the data follows a
normal distribution). The choice of the Shapiro-Wilk test is motivated
by its higher power for a given significance when compared to other
tests.  We first test the AMXPs+NXPs sample (henceforth referred to as
LMXBs) for normality, under the hypothesis that both the AMXPs and
NXPs belong to the same underlying population. We use a significance
level of $\alpha=0.05$ throughout this work, which means that we have
a 5\% chance of false positives for each test performed.  The SW test
gives a p-value of $p<0.02$ and thus we reject our null hypothesis at
the 5\% significance level. A quantile-quantile plot of the sample
shows how the data deviates from the expected $45^{\circ}$ line of a
normal distribution (see Figure~\ref{fig:qqplot}).

\begin{figure}
  \centering
  \rotatebox{-90}{\includegraphics[width=0.8\columnwidth]{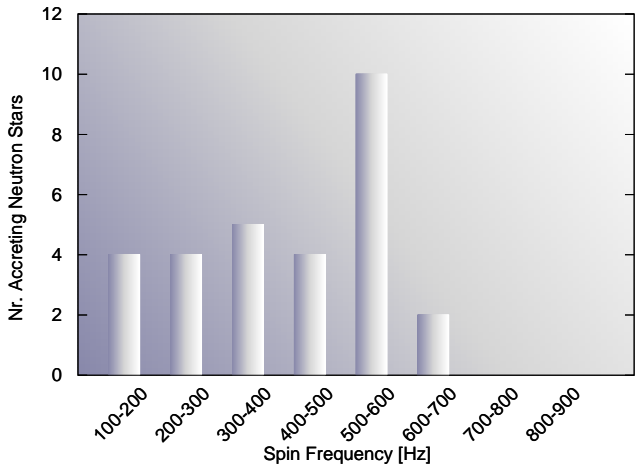}}
  \caption{Histogram of accreting neutron stars comprising all known AMXPs and NXPs (as reported in Table~\ref{tab:sou}). No source is known above $\approx619\,\rm\,Hz$.}
  \label{fig:his}%
\end{figure}

The deviation from normality is evident also from a simple by-eye
inspection of the histogram of the spin distribution shown in
Figure~\ref{fig:his}, which displays a prominent peak around the
500-600 Hz bin (see Table~\ref{tab:sou} for the full list of pulsating
LMXBs used in this work).

\begin{table}
  \caption{Accretion and Nuclear Powered Millisecond Pulsars }
\centering
\begin{tabular}{ccc}
\hline
\hline
Source Name & Spin Frequency [Hz] & Type \\
\hline
4U 1728--34	& 363    & NXP\\
KS 1731--260	& 524     &NXP\\
IGR J17191--2821	& 294	&NXP\\
4U 1702--429 	& 329&NXP\\
\hline
\textbf{SAX J1750.8--2900} & 601     &NXP\\
\textbf{GRS 1741.9--2853} & 589 	&NXP\\
\textbf{EXO 0748--676}	& 552	& NXP\\	
\textbf{4U 1608--52}	& 619    &NXP \\	
\textbf{4U 1636--536}	& 581     &NXP\\
\textbf{MXB 1659--298}	& 567   &NXP\\
\textbf{Aql X--1}          &  550&AMXP\\
\textbf{IGR J00291+5934}   &  599&AMXP\\
\textbf{PSR J1023+0038}	  &  592&AMXP\\
\textbf{XSS J12270--4859}   &  593&AMXP\\
\hline
SAX J1808.4--3658 &  401  &  AMXP\\
XTE J1751--305    &  435 & AMXP\\
XTE J0929--314    &  185&AMXP\\
XTE J807--294     &  190&AMXP\\
XTE J1814--338    &  314&AMXP\\
HETE J1900.1--2455&  377&AMXP\\
Swift J1756.9--258&  182&AMXP\\
SAX J1748.9--2021 &  442&AMXP\\
NGC6440 X-2  	  &  206 &AMXP\\
IGR J17511-3057   &  245 &AMXP\\
Swift J1749.4-2807&  518 &AMXP\\
IGR J17498-2921   &  401 &AMXP\\
IGR J18245-245	  &  254&AMXP\\
MAXI J0911--655    &  340&AMXP\\
IGR J17602--6143   &  164&AMXP\\
\hline
\hline
\end{tabular}\label{tab:sou}
\emph{\footnotesize{\\The sources highlighted in bold are discussed in Section~\ref{sec:sou}.}}
\end{table}

We then perform the SW test for all other pulsar samples, with the
results summarized in Table~\ref{tab:pulsars}. We cannot reject the
null hypothesis only for the spiders and the AMXPs samples.

\subsection{What is the Spin Distribution of Radio Millisecond Pulsars?}

\begin{figure}
  \centering
  \rotatebox{0}{\includegraphics[width=1.0\columnwidth]{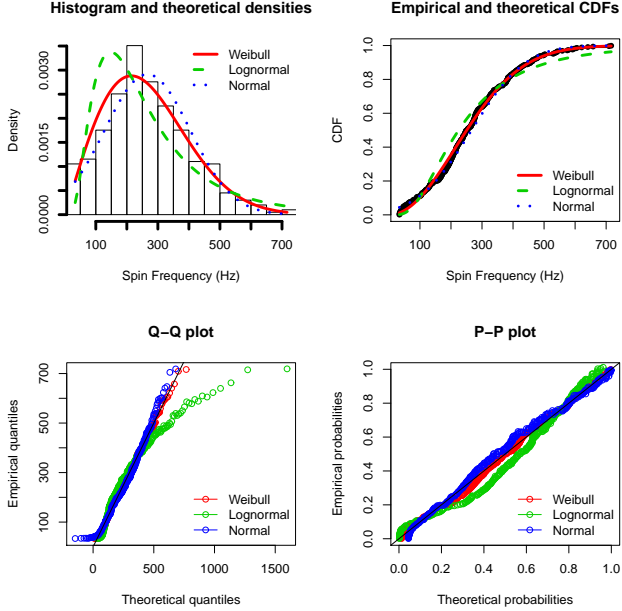}}
  \caption{Probability density of RMSPs (top left panel) with three theoretical distributions fitted (Weibull, lognormal and normal). Top-right panel: empirical and theoretical cumulative density functions for RMSPs. Bottom-left panel: empirical quantile-quantile plot for RMSPs. Note how the data show the strongest deviations from  the theoretical log-normal distribution (oblique solid black line). Bottom-right: empirical probability plot of RMSPs. }
  \label{fig:4pl}%
\end{figure}

\cite{lor15} showed in an analysis of a sample of 56 RMSPs that their
distribution was consistent with being log-normal. \cite{tau12} and
\cite{pap14b} also showed that the distribution of RMSPs is not
consistent with a normal distribution, in line with our previous
findings. Here we try to characterize the spin frequency distribution
of all RMSPs by considering three theoretical distributions:
log-normal, normal and Weibull distribution. The latter has the form:
\begin{equation}
f(x,k,\gamma) = \frac{k}{x}\left(\frac{x}{\gamma}\right)^{k}\,e^{-(x/\gamma)^k}
\end{equation}
where $k$ and $\gamma$ (both assumed positive) are fitting parameters known as shape and scale
parameters, respectively. We fit
the three distributions and then compare the agreement
between the theoretical expectations and the actual data in four plots
in Figure~\ref{fig:4pl}: a histogram of empirical and theoretical
densities, a cumulative distribution function plot, a
quantile-quantile plot and a P-P plot (that compares the empirical
cumulative distribution function of our data with the three
theoretical cumulative distribution functions).  The strongest
deviations are seen for the log-normal distribution, followed by the
normal distribution. The Weibull distribution with shape
$k=2.10\pm0.08$ and spread $\gamma=292\pm8\rm\,Hz$ gives the best match with
the data~\footnote{We note that a Weibull distribution with $k=2$ is equivalent
to the Rayleigh distribution, i.e. a distribution that often emerges
in physical processes that involve scattering or that involve the
sum of two normally distributed and independent vectorial components~\citep{pap02}.}. 

To determine whether the best match of the Weibull distribution 
really is superior to the normal and log-normal distributions we look at
the goodness of fit statistics (package \emph{fitdistrplus}), in
particular we use the Bayesian information criterion
(BIC;~\citealt{sch78}). The BIC is an optimal criterion to select the
best model among many since it introduces a penalty for each
additional free parameter used in the fit. In this way the BIC can
prevent us from erroneously concluding that a model with many free parameters
is necessarily better than a model with fewer free parameters. This  is
 not possible with a simple $\chi^2$ goodness of fit test.

The difference between the minimum BIC number (5036.4 for the Weibull
distribution) and the other two BIC numbers of the normal and
log-normal models is above 30, indicating a very strong evidence
against the higher BIC models~\citep{kas95}. A one-sided
Kolmogorov-Smirnov test  confirms our results, giving a $p$-value of
0.38 for the Weibull distribution and values much smaller than 0.01
for the normal and log-normal distributions.  Again, we caution that
this does not mean that the Weibull is necessarily the true underlying
distribution of RMSPs (we have not considered for example the $\beta$
distribution and other viable options) but that among the models
considered we reject the normal and the log-normal distribution and
suggest a Weibull as having higher likelihood of being the true
underlying distribution.

\begin{figure}
  \centering
  \rotatebox{0}{\includegraphics[width=1.0\columnwidth]{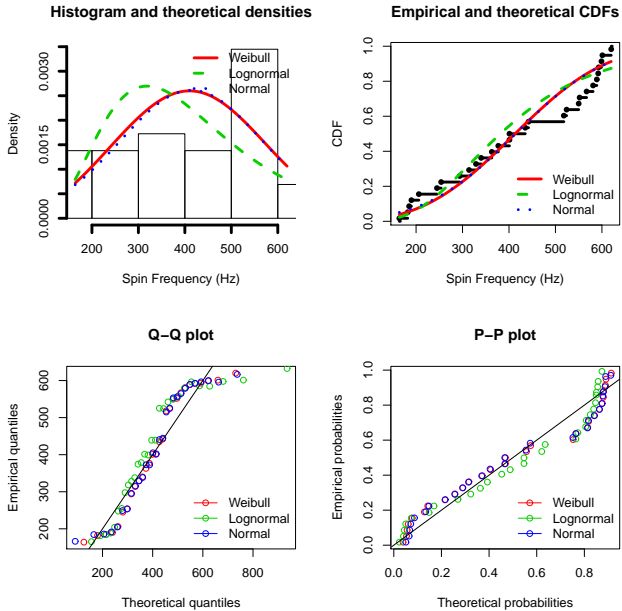}}
  \caption{The same as for Figure~\ref{fig:4pl} but for the LMXB sample. }
  \label{fig:4pllmxbs}%
\end{figure}

A similar analysis applied to the LMXBs shows deviations from all three theoretical distributions (see Figure~\ref{fig:4pllmxbs}). Spiders show a good match both for a normal distribution and for the Weibull, with relatively strong deviations seen in the log-normal model (and BIC number higher than 7 and 9 from the Weibull and normal BIC numbers).

\section{The Spin Distribution of Accreting Neutron Stars}\label{sec:dis}

The separation of fast rotating accreting neutron stars into AMXPs and
NXPs might not be the best way to split the population of LMXBs (see
Table~\ref{tab:sou}) because the presence of accretion powered
pulses -- which distinguishes AMXPs from NXPs, may not  be
related to their spin frequency.

\begin{figure}
  \centering
  \rotatebox{0}{\includegraphics[width=1.0\columnwidth]{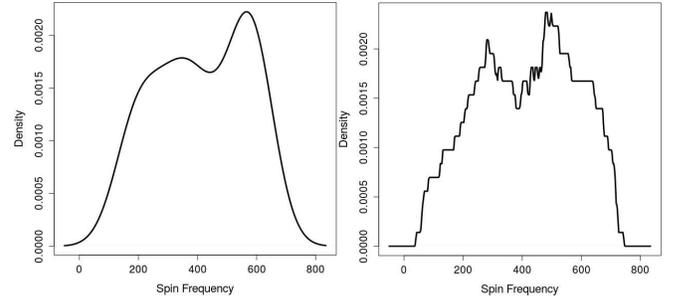}}
  \caption{Kernel Density Estimation for the spin distribution of accreting neutron stars. A Gaussian kernel  has been used in the left panel, whereas a rectangular (and thus discrete) kernel has been used in the right panel. The bandwidth of the kernels is 71 Hz in both cases. The KDE is strongly suggestive of a bimodality in the spin distribution.}
  \label{fig:kde}%
\end{figure}

We thus consider the  population of AMXPs and NXPs together
and then use a kernel density estimator (KDE) to try to extract 
features from the spin distribution. Indeed, although histograms are
the simplest non-parametric density estimators, their properties
heavily depend on the choice of the bin width (beside being discrete
estimators by definition). We use a Gaussian kernel, which is
continuous, and a rectangular discrete kernel. The KDE is performed
with the package \emph{density} and the result is shown in
Figure~\ref{fig:kde}. The distribution seems to be bimodal, with a
clear spike around $\sim550$ Hz and a broader peak around $\sim300$
Hz.

\begin{figure}
  \centering
  \rotatebox{0}{\includegraphics[width=1.0\columnwidth]{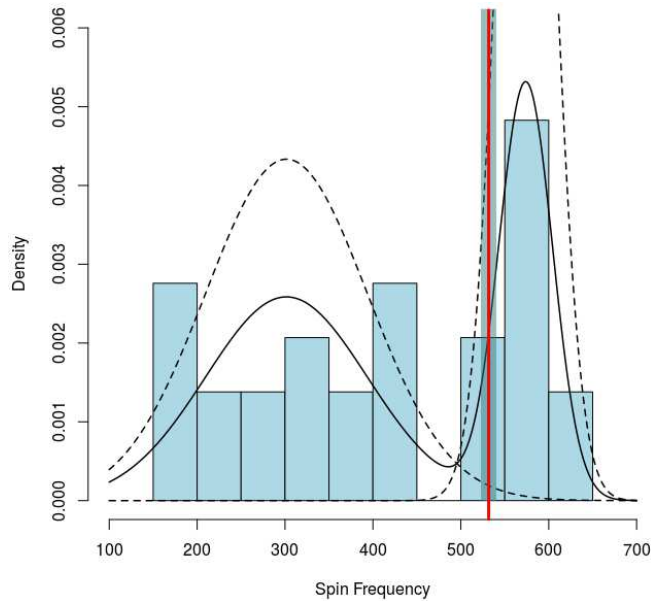}}
  \caption{Mixture model for the spin distribution sample in Table~\ref{tab:sou}. The dashed lines represent the initial distributions $\mathcal{D}_1$ and $\mathcal{D}_2$. The solid black line is the final mixture model after convergence.
    The vertical red line and the shaded gray area around it is the location of the cut-point which divides the sample into two sub-populations. 
    The histogram of the spin distribution is plotted in 10 bins to display the cut-point at 540 Hz more clearly and has no influence on the final results.}
  \label{fig:mix}%
\end{figure}

In summary, the evidence suggests that:
\begin{itemize}
\item our LMXB sample is inconsistent with
being drawn from a population following a normal distribution;
\item a by-eye inspection of the spin distribution shows a prominent peak at 500-600 Hz;
  \item a KDE of the spin distribution appears bimodal.
\end{itemize}
Given these observations, 
we are  motivated to model the data to extract information under
the hypothesis that we are not dealing with a single
population.  In particular, we want to test whether there is evidence
for two sub-populations (not necessarily being AMXPs and NXPs) within
a parent population without specifying which of the 29 data points of
our sample belongs to which sub-population.  To do this we use a fixed
mixture model (e.g.,~\citealt{con00}), which uses an
expectation-maximization algorithm~\citep{guo12} for fitting mixture
models.  A mixture model is a statistical test designed exactly for
this purpose and has the advantage that it can estimate a ``cut-point''
between the two sub-populations. We use a mixture model with
Gaussians, although other choices are possible. The mixture models
operate with two parameters: a location parameter $\mu$, where the data
are concentrated and that corresponds to the mean value of the
distribution (for a Gaussian), and a scale parameter $\sigma$ which
gives the spread of the data and corresponds to the standard deviation
for a Gaussian.  The mixture model then is defined as:
\begin{equation}
  f(\nu|\lambda,\mu_1,\sigma_1,\mu_2,\sigma_2) = \lambda\cdot\mathcal{D}_1(\nu|\mu_1,\sigma_1)+(1-\lambda)\cdot\mathcal{D}_2(\nu|\mu_2,\sigma_2)
\end{equation}
where $\nu$ is the spin frequency, $\lambda$ is the mixture
parameter and $\mathcal{D}_1$ and $\mathcal{D}_2$ are the
distributions used (in this case Gaussians).  The mixture parameter
$\lambda$ acts as a weight for the two distributions
$\mathcal{D}_1$ and $\mathcal{D}_2$.  The function
$f(\nu|\lambda,\mu_1,\sigma_1,\mu_2,\sigma_2)$ is then evaluated
with the expectation-maximization (EM; package \emph{cutoff})
algorithm that is optimal for parameter estimation in probabilistic
models with incomplete data sets.  In short, the EM algorithm works in
two steps, the first is the ``expectation'', where each member of the
sample is assigned a guessed probability to belong to one of the two
sub-populations. In the second step (the 'maximization'), the maximum
likelihood estimation is calculated. The two steps are then repeated
until convergence.  The probability that an accreting neutron star 
belongs to either of the distributions $\mathcal{D}_1$ and $\mathcal{D}_2$ is
calculated as:
\begin{eqnarray}
  p_1 &=& \frac{\lambda\cdot\mathcal{D}_1(\nu|\mu_1,\sigma_1)}{\lambda\cdot\mathcal{D}_1(\nu|\mu_1,\sigma_1)+(1-\lambda)\cdot\mathcal{D}_2(\nu|\mu_2,\sigma_2)}\\\nonumber\\\nonumber\\
  p_2 &=& \frac{(1-\lambda)\cdot\mathcal{D}_2(\nu|\mu_2,\sigma_2)}{\lambda\cdot\mathcal{D}_1(\nu|\mu_1,\sigma_1)+(1-\lambda)\cdot\mathcal{D}_2(\nu|\mu_2,\sigma_2)}
  \end{eqnarray}
We then use our significance level of $\alpha=0.05$ (i.e., the false
alarm probability), along with 20,000 Monte Carlo simulations to
calculate the location of the cut-point that separates the two
distributions and we find a value of about 540 Hz.  The results of our
test for the parameters $\mu_1$, $\mu_2$, $\sigma_1$ and $\sigma_2$,
$\lambda$ and the cut-point are reported in Table~\ref{tab:mm} and
Figure~\ref{fig:mix}.

We then calculate the BIC numbers to estimate the significance for
the presence of two sub-populations rather than a single one (with
the package \emph{mclust}). The BIC number differences are larger than about 7, which implies a strong evidence for the presence of the two
sub-populations\footnote{The BIC numbers can be transformed into a Bayes factor~\citep{wag07}, which, in this case, would correspond to a value of about 40.0.}.
Finally, we use a parametric bootstrap method (package \emph{boot}) to test the
null hypothesis of a one-population spin distribution vs. the two components one.
We produced $10^5$ bootstrap realizations of the likelihood ratio statistics and checked the final $p$-values of the two models. A one-population model gives a $p$-value of less than $1\%$, whereas the two-subpopulations model gives $p\gg5\%$, which  confirms our previous findings. 

\begin{table}[!ht]
\centering
\caption{Mixture Model Parameter Estimation}
\begin{tabular}{lrr}
  \hline\hline
  & Estimate & Confidence Interval (95\%)\\
  \hline  
$\mu_1$ (Hz)& 302 & 255--348 \\
$\sigma_1$ (Hz) & 92 & 68--135\\
$\mu_2$ (Hz) & 574 & 555--593\\
  $\sigma_2$ (Hz) & 30 & 21--48\\
  $\lambda$ & 0.6 & 0.4--0.8\\
\textbf{Cut-point (Hz)} & \textbf{538} & \textbf{526--548}\\
\hline
\hline
\end{tabular}
\emph{\footnotesize{}}
\label{tab:mm}
\end{table}

\section{Is there a Common Underlying Pulsar Population?}\label{sec:pop}

If we consider our two sub-populations, then the slower systems have
an average spin frequency of about $\langle\nu\rangle\approx302\pm92$ Hz (see Table~\ref{tab:mm}). 
This is within one standard deviation from the average RMSP spin frequency
which is $\approx250$ Hz (see also~\citealt{hes08} and \citealt{pap14b}).
Therefore, the suggestion that the slow sub-population is indeed the progenitor
population of RMSPs is consistent with observations.

This leads us to consider whether the samples of AMXPs, NXPs, RMSPs and spiders
are compatible with being drawn from the same underlying
population. To check this we use a $k$-samples Anderson-Darling test
(package \emph{kSamples}), which does not  require us to specify the
distribution function of the population (which, in our case, is indeed
unknown). We first compare LMXBs and RMSPs: the $k$-samples AD test
gives a $p$-value of about $\approx10^{-7}$ ($H_0$: both samples are
drawn from the same underlying distribution), and thus we reject the
null hypothesis that both samples come from a common population.
Similar results are obtained if we compare spiders and RMSPs
($p$-value $\approx10^{-8}$) and spiders vs. NXPs ($p$-value
0.005). The comparison between AMXPs and NXPs ($p$-value 0.038) shows
instead a marginal difference.  The only comparison where $H_0$ is
clearly not rejected is the one between spiders and AMXPs ($p$-value
0.73). The results are summarized in Table~\ref{tab:adk}.
\begin{table}
  \caption{Anderson-Darling $k$-Sample Test. $H_0$ is the null hypothesis that both samples are drawn from the same underlying distribution.}
\centering
\begin{tabular}{lll}
\hline
\hline
Samples & $p$-value & Reject $H_0$\\
\hline

LMXBs \& RMSPs & 2.3e-8 & yes\\
Spiders \& RMSPs & 5.5e-8& yes \\
Spiders \& NXPs & 0.006 & yes \\
AMXPs \& NXPs & 0.038 & marginal \\
Spiders \& AMXPs & 0.73 & no\\
\hline
\end{tabular}\label{tab:adk}\\
\end{table}

Since both the SW and the $k$-sample Anderson-Darling test show no detectable difference
in the distributions of the AMXPs and spiders we check whether the
reason might be due to the low power of our experiment.
The power of the experiment is  
an estimate of the probability to find a real effect 
and can be defined as one minus the probability of
a Type II error (also called a false negative).
For this we need to first estimate the effect size, i.e., the
quantitative difference between the two samples. 
We do this by using Hedge's $g$~\citep{hed81}, which is an appropriate
estimator when the sample sizes are not equal, under the hypothesis that
both samples are drawn from a normal distribution. The $g$ value is 0.1
which indicates a small effect (if any). The 95\% confidence interval
of $g$ is $[-0.43, 0.64]$, which means that if there's "an effect", i.e. a 
difference between the two distributions, then it is only slightly more likely
to have a larger mean spin frequency for the AMXPs over the spiders. 
The power of our experiment is also very small, of the order of 10\% and
therefore it is not surprising that we detect no significant difference
between the two samples. We stress that this does not necessarily imply 
that AMXPs and spiders come from a single population. A more
conservative conclusion would be that the sample sizes are too small to
detect a (presumable relatively small) difference.

\subsection{Selection Effect of Small Sample Size}

We carried out a set of Monte-Carlo simulations to verify whether the
spin-down of the RMSPs due to magnetic dipole radiation can explain
the average spin difference with the slow LMXB sub-population.  First
we selected all spin-down values of known RMSPs from the ATNF
catalogue\footnote{atnf.csiro.au/people/pulsar/psrcat/}. Then we
remove all sources that belong to globular clusters (since their
spin-down measurement is affected by the cluster gravitational
potential well).  We also remove all sources which belong to the
population of spiders as defined in the preceding sections. We 
simulate $10^5$ pulsars with a spin frequency randomly selected from a
distribution equal to the one found in this work (Weibull with $k=2$
and $\gamma=292$ Hz). We then simulate $10^5$ ages ($\tau_{\rm age}$)
following a uniform distribution with boundaries $10^8$ and $10^{10}$
yr and spin frequency derivatives ($\dot{\nu}$) with the same
distribution as the observed ones. We finally randomly select a
pulsar, spin-frequency derivative and age and increase the pulsar
frequency as: $\nu_{\rm final} =\nu_{\rm initial}+\dot{\nu}\,\tau_{\rm
  age}$.  Then we calculate the mean, median and standard deviation of
our evolved pulsar population. Of course, the results are sensitive
to the maximum allowed spin frequency, so we select three cases, one
with $\nu_{\rm max}=730$ Hz (compatible with the cutoff suggested
by~\citealt{cha03}), another with $\nu_{\rm max}=1000$ Hz and finally
one with $\nu_{\rm max}=1500$ Hz. When $\nu>\nu_{\rm max}$ we remove
our pulsar from the population. We then sample our simulated
population by extracting 19 randomly selected objects (since this is
the number of known AMXPs) and calculate the mean, median and standard
deviation of this random sample. We repeat the procedure $10^3$
times and for each run we perform a Shapiro-Wilk test. Our results
show that when $\nu_{\rm max}\approx700-1000$ Hz, our final sample is
consistent with a normal distribution in approximately 9/10 of the
cases, whereas increasing the spin-limit to 1500 Hz makes the distribution consistent in just
3/10 of the cases.  Therefore, although we cannot exclude that a spin
cutoff is present at sub-ms periods, our simulations favor a lower
cutoff. This is, again, compatible with the current observations.

\section{Observations vs theory}\label{sec:obstheory}

Our analysis suggests the presence of two sub-populations of neutron stars in LMXBs, separated at
around 540 Hz, with the fast population having a very narrow standard
deviation of only 30 Hz. Indeed, a peak due to clustering of several
sources at around 500-600 Hz is already prominent with a by-eye
inspection of the spin distribution, see Figure~\ref{fig:his}. Having drawn this conclusion, we are led to the obvious question: what is the mechanism that leads to the clustering of this fastest spinning objects? The issue of a ``speed limit'' for accreting neutron stars is not new. However, it is usually discussed in the context of the spin-equilibrium associated with accretion and whether additional torques (in particular, due to gravitational waves) are required to explain the data. The usual evidence  in favor of such a mechanism is the absence of neutron stars spinning near the break-up limit. Gravitational-wave emission provides a natural explanation as it does not require particular fine-tuning of the accretion flow and also has no direct connection to the star's magnetic field \citep{pp78, wag84}. The evidence of a clustering at the fastest observed spins provides additional arguments in this direction. Whatever the mechanism is that causes the clustering, it must set in sharply as soon as the stars reach a given spin rate.  Again, gravitational wave torques, which scale with a high power of the spin frequency ($\nu^5$ for deformed stars) may lead to exactly this behaviour. However, this is a phenomenological argument. It is entirely possible that the answer has nothing whatsoever to do with gravitational waves. The accretion torque may simply become much less efficient as soon as the star reaches above 540~Hz.

These issues have been discussed at length in the literature \citep{cha03, pat10, pat12a} and we will not be able to resolve them here. However, we can phrase the questions in a new light. As we are arguing in favor of a distinct population of fast spinning LMXBs, we can ask whether the systems that belong to this population are somehow different from their slower spinning relatives. Are there other observed phenomena that make these systems distinct? If so, do these observations provide a clue to the underlying explanation?

Additional evidence in support of the conclusion that the observed peak at 500-600 Hz is
connected to some underlying physical phenomenon and  not due to
chance may come from the thermal emission.  If we consider all 20 neutron star LMXBs with both known
spin frequency and a measured surface temperature, then all sources 
falling in the fast sub-population have a measured $T$ while the remaining ones have only upper limits (with the exception of one source plus another marginally outside the 540 Hz cut-point confidence interval \citep{ho11, has12, gus14}. This may be a fairly weak argument, as the 
 temperature upper limits are often not very constraining, but the fact that all sources in the fast population have a measured temperature might indicate that they are hotter, on average, than those in the slow population.
If this is, indeed, the case then any scenario that explains the spin-distribution must involve some additional heating. As we will see in the following, an unstable r-mode large enough to balance the spin-up torque due to accretion would naturally re-heat the star to these levels, and could provide an explanation. Nevertheless, several aspects of this picture remain problematic.

\subsection{The Accretion Torque}\label{sec:acc}

As a starting point for a more detailed discussion, it is natural to consider the accretion torque.
All available accretion models build on the idea that the flow of accreting matter will, at some point become dominated by the star's magnetic field. 
The material torque in this interaction region
is usually approximated by
\begin{equation}
\dot J = \dot{M} \sqrt{GM R_M} = N_\mathrm{a} 
\label{mat1}\end{equation} 
where the magnetosphere radius is given by
\begin{equation}
R_{{M}}= \left({\mu^4\over 2GM\dot{M}^2} \right)^{1/7}
\end{equation}
with  $\mu= BR^3$ the star's magnetic moment, $M$ the neutron star's mass and $R$ its radius. The actual location of the transition to magnetically dominated flow is obviously not this precise, so it is common to introduce a factor $\xi \approx 0.1-1$ to parametrize the unknowns. We will,however, omit this factor here in the interest of simplicity. 

Meanwhile, outside the co-rotation radius
\begin{equation}
R_c= \left( {GM\over \Omega^2}  \right)^{1/3}
\end{equation}
where $\Omega$ is the angular spin frequency of the star,
the magnetic field lines rotate faster than the local Keplerian speed, resulting in a negative torque. 
If $R_M>R_c$ the accretion flow will be centrifugally inhibited and matter may be 
ejected from the system\footnote{\cite{spr93} showed that material can be expelled from the disk only if $R_M\gtrsim\,1.3R_c$. However, this has no influence on the arguments presented in our discussion and we will thus ignore this difference for the sake of simplicity.}. It is easy to see that this will happen 
if the spin period becomes very short, or the rate of flux of material
onto the magnetosphere drops. This is the
propeller regime. In this phase accreting matter is flung away
from the star, leading to a spin-down torque. In order to account for this effect we alter the material torque \citep{ho14} \footnote{This model is somewhat simplistic, but it captures the main features of the problem. Other prescriptions tend to be more complex, but would lead to broadly the same conclusions.}:
\begin{equation}
 {\dot{J}}  
=   N_\mathrm{a}  \left[ 1 - \left({R_M \over R_c} \right)^{3/2} \right]
 =  N_\mathrm{a} ( 1 - \omega_s)
\label{mat2}\end{equation}
where we have introduced the so-called fastness parameter   $\omega_s$. Within this model, it is easy to see that the system will not spin up beyond
 $\omega_s = 1 \longrightarrow R_M = R_c$. In essence, we can estimate (for given stellar magnetic field, accretion rate etcetera) at which point the system reaches spin-equilibrium. Broadly speaking, this accretion model will lead to a flat spin-distribution unless we fine-tune the magnetic fields. Without adding features to the model, it is certainly difficult to explain the two populations from Figure~\ref{fig:mix}.

If we observe the star to spin-up during accretion, then we can also use these results to constrain the magnetic field. In order to link the observed X-ray flux to the accretion rate, one would typically take
\begin{equation}
L_X = \eta \left( {GM \over R} \right) \dot M 
\end{equation}
where $\eta$ is an unknown efficiency factor, usually taken to be unity. 

As we progress, it is useful to consider the uncertainties of different torque models. The main uncertainties are well known, and relate to three
factors:
\begin{itemize}
\item[1.] the nature of  the accretion flow (thin or thick) in the  region close to the neutron star magnetosphere,  
\item[2.] the disk/magnetosphere interaction (typically encoded in the $\xi$ factor), 
\item[3.] the estimate of the mass accretion rate due to  transient behavior and the presence of outflows (leading to additional torques, which tend to be equally uncertain). 
\end{itemize}
In a recent work, \cite{dan16} reviewed these uncertainties and concluded that the spin equilibrium of
AMXPs can be explained without the need of additional spin-down
mechanisms like gravitational waves.  However, these models cannot naturally explain the pile-up of sources between $\approx 500-600$ Hz, and the conclusion only
applies to sources with a dynamically important magnetosphere. As we will see later, it is not clear that all LMXBs fall in this category. 
If the
magnetosphere is too weak (or even absent) to truncate the accretion
disk then an additional spin down mechanism will definitely be necessary.

In addition, \cite{spr93} and then \cite{dan10,dan12}
showed that the presence of a magnetosphere can alter the regions of
the accretion flow close to the co-rotation radius. The plasma
interacting with the magnetosphere absorbs part of the neutron star
angular momentum and remains temporarily close to the co-rotation
radius altering the density/temperature profile of the inner disk regions. If such a
\emph{trapped disk} configuration really exists in accreting neutron
stars is not yet clear, but there is  evidence 
that it might be a viable option~\citep{pat09c,pat13c,jao16,vde17}.

The main difference between a trapped disk and the propeller model were discussed 
in \citet{dan16}. In essence, the trapped disk model predicts a stronger spin-down 
than propeller during the declining portion of an outburst which can lead to a 
lower spin frequency limit. This is due to the fact that the strength of the spin-down
in the propeller model scales with the amount of mass available in the inner disk regions, 
which drops substantially during the declining outburst phase. 
If a trapped disk is formed in most systems (with a magnetosphere) then
the net neutron star spin-up during the outburst may be strongly reduced. Moreover, if a residual
trapped disk persists even during quiescence then there is an additional spin down torque and the
discrepancy between the average spin frequency of accreting neutron stars
and RMSPs could be reconciled. However, this involves two relatively
strong assumptions: the formation of a trapped disk and the
persistence of this disk once the source accretes at very low mass
accretion rates. Although this is certainly possible (e.g., \citealt{dan10} show that 
once a trapped disk is formed it is basically impossible to "untrap" it), this scenario
awaits further observational evidence. More detailed calculations are also needed to verify whether 
a trapped disk can truly explain the cutoff at high frequencies and the presence of two 
sub-populations of LMXBs.

Another potential problem of the trapped disk is the fact that, in at least three AMXPs ~\citep{pat12, pat10, har11, pap11, rig11} a spin-down has been measured during quiescence and
its value is entirely consistent with the expected value from a magnetic dipole spin-down of a neutron star with a field of the order of $10^8$ G. There is  no clear evidence for
enhanced spin-down (at least during quiescence) in AMXPs. Moreover, it is difficult to observe spin-down during the declining portion of an outburst due to the short duration of this phase and/or the lower quality of the data. The current upper limits on the spin-down are thus still compatible with both a trapped disk and a propeller model (see, for example, \citealt{pat16} for a discussion of the system SAX J1808.4--3658).

\subsection{Gravitational waves}\label{sec:gws}

Let us now consider the conditions for gravitational waves to be effectively spinning down accreting neutron stars. 
Most importantly, the asymmetry of the accretion flow, with matter channeled onto the magnetic poles of the star, 
should induce some level of quadrupole asymmetry in the stars moment of inertia tensor. Quantifying this in terms of an ellipticity, $\varepsilon$, it is easy to show that the associated spin-down torque will balance the accretion spin-up (given by $N_\mathrm{a}$ for simplicity) if
\begin{equation}
\varepsilon \approx 7\times 10^{-9} \left( {\dot M \over 10^{-9} M_\odot/\mathrm{yr}}\right)^{1/2} \left( {600\ \mathrm{Hz} \over \nu}\right)^{5/2}
\end{equation}
Given this estimate, it is wroth making two observations. First of all, the required $\varepsilon$ is smaller than the deformation required to break the crust \citep{has06,jo13} so there is no reason to rule out this scenario. Second, the required deformation is broadly in line with estimates for the quadrupolar temperature/composition asymmetry induced by  pycnonuclear reactions in the
deep-crust \citep{bil98,ush00}. We should also keep in mind that the accretion torque (Eq.~\ref{mat2}) would be weaker, so the actual ellipticity required could well be smaller. 

Another possibility is that the gravitational-wave driven instability of the inertial r-modes \citep{and98,owen} is active in these systems. Based on the current estimates for the damping mechanisms (shear- and bulk viscosity, viscous boundary layers, superfluid vortex mutual friction etcetera), this seems plausible (see \citealt{hasrmode} for a recent review). In fact,  the estimated temperatures for the LMXBs would put a number of known systems in the unstable regime \citep{ho11, has12}. Moreover, the r-mode instability would readily balance the accretion torque. Quantifying this statement in terms of the mode amplitude $\alpha$, we need 
\begin{equation}
\alpha \approx 6\times 10^{-6} \left( {\nu \over 600~\mathrm{Hz}}\right)^{-7/2}  \left( {\dot M \over 10^{-8} M_\odot/\mathrm{yr}}\right)^{1/2} 
\end{equation}
As this is much smaller than the expected saturation amplitude for the instability (due to the nonlinear coupling with shorter wavelength modes), $\alpha_\mathrm{sat} \sim 10^{-3}$ \citep{arr03, bond07}, the gravitational waves from r-modes should be able to prevent these systems from spinning up. However, this explanation is problematic.  In particular, the fact that the instability is expected to grow much larger than what is required to overcome the accretion torque means that neutron  stars should not be able to venture far into the instability region. This means that the actual instability limit should be such that all known systems are (at least marginally) stable. However, the current theory does not provide a mechanism that predicts this (at least not without fine-tuning, see e.g. \citealt{gus14, chug17}). The internal physics is, of course, complex and we may simply be missing something. It may be worth noting that the typical instability curve shifts by a few tens of Hz if we vary the neutron star mass from $1.4M_\odot$ to $2M_\odot$. This may be a complete coincidence but there could be a connection with the width of the distribution of fast spinning LMXBs. 
Furthermore an r-mode growing to amplitudes large enough to affect the spin-evolution of the system would also reheat the star, leading to an internal temperature of \citep{has12}:
\begin{equation}
{T}\approx 10^8 \mbox{K} \left[\left(\frac{\alpha}{10^{-6}}\right)\left(\frac{\nu}{600 \mbox{Hz}}\right)\right]^{1/5}.
\end{equation}
It is thus possible that the faster, and hotter, systems, could harbor an r-mode large enough to significantly affect the spin-evolution and halt the spin-up of the star \citep{ho11, mah13}.

As gravitational waves may well be  emitted by accreting neutron
stars, it is natural to consider the different mechanisms that may be responsible
for the generation of the waves and what constraints  we can place on
these mechanisms from the observed spin distribution. The fact that the pile-up at higher frequencies seen in Figure~\ref{fig:his} would be naturally explained by gravitational waves becoming relevant at a critical rotation rate,  provides strong motivation for renewed efforts in this direction.

\section{Individual Sources}\label{sec:sou}

Returning to observations, let us ask what
type of sources belong to the peaked 
sub-population, i.e. the fastest rotators.  Thus, we identify ten sources,
four of which are AMXPs and six NXPs (highlighted in bold in Table~\ref{tab:sou}).
Although the spin evolution of NXPs is not known, the fact that we
have some AMXPs for which the spin evolution is constrained to a
certain degree allows us to identify possible anomalies that might
hint at the presence of a common effect, which is triggered (or becomes more prominent)
once the neutron stars pass the 540 Hz cut-point.

%

\subsection{Aql X--1}

Turning to individual systems, it is perhaps natural to start with Aql X--1. This is one the most mysterious accreting neutron stars because it
has been observed to pulsate for just 150 seconds out of a total
observed time of almost 2 Ms~\citep{cas08}. \cite{mes15} showed that
the pulse fractional semi-amplitude went from $\approx6.5\%$ (during the
150 s of pulsed episode) down to
less than $0.26\%$ in the same outburst, with similar order of
magnitude upper limits for the other 18 outbursts analyzed.  Such an
incredibly low duty cycle for the presence of pulsations has spurred a
number of hypotheses on their origin, with
no model currently being able to explain the observations (see \citealt{cas08} and \citealt{mes15} for discussion).
\cite{mes15} proposed that the only plausible model is one in which a
mode of oscillation, with azimuthal number $m$ and
frequency $\nu_{\rm mod}{\sim}10^{-2}-1$ Hz, is triggered. However, oscillation
modes need to be excited by some event and so far there is no
evidence for  a trigger shortly before the beginning of the
pulsing episode. In addition, one would have to understand why any excitation mechanism would single out such a specific low-frequency mode.

\subsection{IGR J00291+5934}

In the case of IGR~J00291+5934 a strong spin-up during an outburst of accretion was observed to be followed by a slow drop off in the spin-rate \citep{pat10, har11, pap11}. This behavior would be naturally explained in terms of a dramatic rise in the accretion torque during the  outburst and standard magnetic dipole spin-down in between outbursts. This is interesting because it allows us to constrain the star's magnetic field in two independent ways. In quiescence, the magnetic field follows from the standard pulsar dipole formula, while the spin up torque encodes the magnetic field through the magnetosphere radius, as in Eq.~\eqref{mat2}.

At first glance, the observed spin-evolution of  IGR~J00291+5934 is consistent with the theory~\citep{pat10}. The magnetic field estimated from the spin down is in line with the maximum magnetic field inferred from the spin-up phase. However, this argument is not quite consistent~\citep{and14}. Basically, the latter estimate is based on setting $\omega_s=1$. This does indeed lead to an upper limit on the magnetic field consistent with the system spinning up, but as this is also the condition for spin-equilibrium the actual predicted spin-up rate would be low. In order to quantify the discrepancy, without changing the torque prescription, we can ask how the result changes if we vary the parameters $\xi$ and $\eta$. The first of these is not useful, as any value of $\xi<1$  leads to a weaker spin-up torque.  Changing $\eta$ is more promising as we end up with a larger $\dot M$ meaning that the torque is enhanced, but this resolution is quantitatively uncomfortable. We need roughly $\eta\approx 0.1$ in order for the magnetic field estimates to be consistent.

It is also important to note, given the context of this discussion, than an additional gravitational-wave spin-down torque would only enhance the problem.

\subsection{XTE~J1751$-$305 \& 4U 1636$-$536}

XTE~J1751$-$305 is a slower spinning system (at 435~Hz, well below our cut-point) which exhibits a similar spin-up/down behaviour to IGR~J00291+5934. In this case the inferred magnetic field  also inconsistent~\citep{and14}, but slightly less so. In this case, we need $\eta\approx 0.25$ in order to reconcile the data. This system is interesting for another reason. \citet{str14} reported a coherent quasi-periodic oscillation during the discovery outburst. They suggested that the frequency of this feature, roughly $0.57 \nu$, would be consistent with the star's quadrupole r-mode. This is an interesting suggestion, given the possibility that these modes may be unstable in fast spinning stars. However, the idea is also problematic. First of all, one would need to explain why it is the rotating frame frequency that is observed rather than the inertial frame one (as one would expect). An argument in favor of this has been put forward by \citet{lee14}. Even so, after confirming that the observed frequency would indeed be consistent with an r-mode (after including relativistic effects and the rotational deformation) \citet{and14} point out that the gravitational waves associated with an r-mode excited to the observed level would spin the star down very efficiently. This would prevent the system from undergoing the observed spin up.

A similar feature was later found in the faster spinning system 4U 1636$-$536~\citep{str14b}, which is above our cut-point. In this case the observed feature was found at $1.44\nu$. This would be roughly consistent with an r-mode in the inertial frame. However, the same provisos apply to this observation. 

\subsection{PSR J1023+0038 \& XSS J12270--4859}

PSR J1023+0038 \& XSS J12270--4859 are two of the three known
tMSPs  (see~\citealt{arc09} and
\citealt{bas14}). PSR J1023+0038, in particular, has been
monitored in radio and X-rays for the past 10 years (e.g., \citealt{pat14}, \citealt{bog14} and \citealt{sta14}).   Spin down
has been measured both in radio~\citep{arc13} and during the accretion
powered phase~\citep{jao16}. The spin down during the latter phase is
particularly difficult to explain with the current accretion torque models (see~\citealt{jao16}
for an extended discussion). The problem is that the system spins down faster when it accretes than it does in quiescence. This led  \cite{has17} to suggest  that
gravitational waves are playing a role in spinning down the neutron star. The quadrupole deformation required to explain the observation is in line with theoretical expectations ($\varepsilon\approx\,5\times10^{-10}$), making the explanation consistent. Moreover, if it is correct then we would have a first handle on what level of accretion induced asymmetries to expect in LMXB. Even if the gravitational-wave signal itself would be too weak to be detectable (given current technology) it would help us model the general population better.

\section{Is There a Magnetosphere?}\label{sec:mag}

Another piece of observational evidence comes from the fact that some of 
these systems might, in fact, lack a magnetosphere strong enough to affect
the dynamics of the plasma in the accretion disk.  Recent searches for pulsations in several LMXBs
have found no evidence for continuous accretion-powered pulsations, with
upper limits on the pulsed fraction of $\approx0.1-0.3\%$ (\citealt{mes15}, Patruno,
Messenger \& Wette 2017, in prep) and down to a few \% fractional
amplitude for brief pulsation episodes of short duration (0.25 s--64
s; Algera \& Patruno 2017, in prep.). The sample analyzed includes
both NXPs and other LMXBs. It is difficult to reconcile such exquisite uniformity of the neutron
star surface with the presence of a strong 
magnetosphere.  

However, in the context of the present discussion this is problematic. If no magnetosphere (or a very
weak one) is present, then no centrifugal barrier can act to mitigate
the spin-up due to the transfer of angular momentum from the accreting
plasma. Basically, we have to use the torque $N_\mathrm{a}$ with $R_M=R$. Then the change in spin
 scales linearly with the amount of mass transferred and
therefore the neutron star keeps spinning up  (not even spinning down during quiescence, since the pulsar spin down
mechanism from magnetic dipole radiation is also halted). Moreover, if there is no magnetosphere, then the accretion flow does not lead to asymmetries on the star's surface and therefore one may not expect the system to develop the deformation required for gravitational-wave emission (although mode instabilities can obviously still play a role and frozen-in compositional asymmetries may still lead to the neutron star being deformed). 

If one accepts that the current upper limits imply the lack of a magnetosphere in these systems then, given that the non pulsating
LMXBs  constitute the majority of accreting neutron stars, there should exist 
very fast spinning ($\nu\gg619$ Hz)   objects which, so far, have not been observed.

There are two, fairly natural, possibilities. First, it could be the case that the magnetic field of NXPs has decayed (for example,
through Ohmic dissipation). If this is the case, then the field will not re-emerge when accretion stops. If this is the case, 
 the NXPs can not be the progenitors of the RMSPs. Instead, there may
 exist an unseen population of very fast rotating neutron stars with
no significant magnetic field. These would only be visible during the accretion phase,
when burst oscillations are produced on their surface. This might explain
why AMXPs and NXPs behave differently and why we do not see any excess
in the spin distribution of RMSPs at high frequencies (whereas we see
a second sub-population among the accreting systems). 
 
Alternatively, the absence of a magnetosphere may indicate that the magnetic field is buried by the accretion flow. The problem of magnetic field burial has been considered in detail for young neutron stars, following supernova fall-back accretion (e.g., \citealt{vig13}) and in some simplified form also for accreting neutron stars \citep{cum01,cum08}. To get a first idea whether this idea is viable for the much lower accretion rates we are interested in we can adapt the usual argument. 

Taking the work by \citet{gep99} as our starting point, we first consider the depth at which the field would be buried. To estimate this we balance the timescale associate with the inflowing matter to that of the Ohmic dissipation. We first of all have
\begin{equation}
t_\mathrm{flow} = {L\over v_r}
\end{equation}
where $L$ is a typical length-scale of the problem, and in the case of accretion we have
\begin{equation}
v_r = {\dot M \over 4\pi r^2 \rho}
\end{equation}
Secondly we need 
\begin{equation}
t_\mathrm{Ohm} = {4\pi \sigma L^2 \over c^2}
\end{equation}
where $\sigma$ is the conductivity.
If $t_\mathrm{flow}<t_\mathrm{Ohm}$ then the magnetic field is frozen in the inward flowing matter. The matter piles up faster than the field can diffuse out and hence we have burial. If the accretion stops, the field emerges on the $t_\mathrm{Ohm}$ timescale associated with the burial depth. 

As we do not expect the field to be buried deep, we consider the envelope where the ions are liquid (the electrons are degenerate and relativistic). Then we have \citep{gep99}
\begin{equation}
\sigma \approx 9\times 10^{21} \left( {\rho_6 \over AZ^2} \right)^{1/3}\ \mathrm{s}^{-1}
\end{equation}
with $\rho_6 = \rho/(10^6$~g/cm$^3$)
which leads to 
\begin{equation}
{t_\mathrm{Ohm} \over t_\mathrm{flow}} \approx  6\times 10^{5} {L_5 \dot M/\dot M_\mathrm{Edd} \over r_6^2 \rho_6^{2/3} (AZ^2)^{1/3}}
\end{equation}
where $\dot M_\mathrm{Edd} = 10^{-8} M_\odot$/yr, $A$ is the mass number of the nuclei and $Z$ is the proton number. As we are interested in the outer region it makes sense to consider Fe$^{56}$, with $A=56$ and $Z=26$. We can also set $r_6\approx 1$, as we are near the star's surface. Then we have
\begin{equation}
{t_\mathrm{Ohm} \over t_\mathrm{flow}} \approx  2\times 10^{4} {L_5 \over \rho_6^{2/3}} {\dot M \over \dot M_\mathrm{Edd} }
\end{equation}
where $L_5=L/10^5$~cm.

We also need to know how the density increases with depth. As we only want rough estimates we  use the pressure scale height
\begin{equation}
H = {p\over \rho g}
\end{equation}
where the gravitational acceleration
\begin{equation}
g= {GM\over R^2} 
\end{equation}
can be taken as constant. From \citet{bro98} we take
\begin{equation}
H \approx 265 \left( {2Z \over A} \right)^{4/3} \rho_6^{1/3} \ \mathrm{cm}  \longrightarrow \rho_6^{2/3} \approx 2\times 10^{5} H_5^2 \end{equation}
with $H_5=H/10^5$~cm.
Finally, it makes sense to let $L\approx H$, so we are left with
\begin{equation}
{t_\mathrm{Ohm} \over t_\mathrm{flow}} \approx  0.1 H_5^{-1}  {\dot M \over \dot M_\mathrm{Edd} }
\end{equation}
and we learn that the magnetic field is buried up to a density  
\begin{equation}
\rho_\mathrm{burial} \approx 7\times 10^{10} \left(  {\dot M \over \dot M_\mathrm{Edd} } \right)^3 \ \mathrm{g/cm}^3
\end{equation}
This estimate  agrees quite well with (extrapolations from) for example \citet{gep99}. 

We also have
\begin{equation}
L_\mathrm{burial} \approx H_\mathrm{burial} \approx 8\times 10^3  \left(  {\dot M \over \dot M_\mathrm{Edd} } \right) \ \mathrm{cm}
\end{equation}
which means that,
once accretion stops, the field will re-emerge after
\begin{equation}
t_\mathrm{Ohm} \approx 10^{10}  \left(  {\dot M \over \dot M_\mathrm{Edd} } \right)^2 \ \mathrm{s}
\end{equation}
That is, the field will emerge  a few hundred years after a star accreting at Eddington goes into quiescence.

Would also need know how long it takes to bury the field in the first place. Somewhat simplistically, this follows from the accreted mass corresponding to the burial estimated depth. This is roughly given by
\begin{equation}
\Delta M \approx 4\pi \rho_\mathrm{burial} R^2 H \approx 4\times 10^{-6}   \left(  {\dot M \over \dot M_\mathrm{Edd} } \right)^{4/3} M_\odot
\end{equation}
and we see that it would also take a few hundred years to bury the field at the Eddington accretion rate. 
Lower accretion rates, such as those of many of the systems we have considered, would lead to a more shallow burial, leading to the field being buried and re-emerging on much shorter timescales. 

These estimates obviously come with a number of caveats. A number of complicating factors may come into play, like possible plasma instabilities \citep{muk13} and the tension of the internal magnetic field, which can lead to sharp gradients and reduce the typical length-scale $L$, thus reducing the amount of mass and the timescale needed for burial \citep{pay04,pay07}. At least at the the back-of-the-envelope level, however, we can see no reason why the temporary burial scenario would not work. A modest level of accretion may lead to a shallow field burial, with the magnetic re-emerging shortly after a system goes into quiescence. 
Furthermore the deformed magnetic field due to accretion may lead to quadrupolar asymmetries and gravitational wave emission, which, depending on the strength of the magnetic field and the degree of burial, may be large enough to contribute to the spin evolution of the system \citep{mp05, pri11}.

In order to improve on these rough estimates, we may want to consider the possibility that the absence of pulsations is due to local field burial in the polar regions. This would perhaps not affect the burial depth very much but we would not need accrete as much mass as we have estimated. Moreover, within such a scenario it could be that there is still a magnetosphere-disk interaction, which could influence the accretion torque. 

\section{Conclusions}

In this paper we have studied the spin distribution of accreting
neutron stars and radio millisecond pulsars. Our analysis (Sections~\ref{sec:cri},~\ref{sec:dis} and~\ref{sec:pop}) shows that
the spin distribution of accreting neutron stars can be best described
by two sub-populations, one at relatively low frequencies with mean
spin frequency $\mu_1\approx300$ Hz and a fast one with $\mu_2\approx
575$ Hz. The fast population is strongly peaked within a very narrow
range of frequencies ($\sigma_2\approx30$ Hz). The two sub-populations
are split at around 540 Hz. The fast sub-population is composed by a
mixture of neutron stars with a magnetosphere and others that
have shown so far only burst oscillations.  

We have shown how all objects in the fast LMXB population have a measured surface temperature, which suggests they might be hotter, on average, than the slow-population
(Section~\ref{sec:obstheory}). 
We have discussed various accretion torque models (Section~\ref{sec:acc}) and
argue that, even when considering the various uncertainties that plague them, 
 no model can naturally explain the presence of a fast sub-population. 
We therefore considered the role that gravitational waves might have 
on the neutron star spin if they are efficiently emitted above the
cut-point at $\approx540$ Hz (Section~\ref{sec:gws}). We find that different lines of evidence suggest that gravitational 
waves might be playing an important role to regulate the spin of accreting neutron stars, although 
we caution that open questions remain on the exact emission mechanism. 
In particular, some sources that fall in the fast sub-population have a behavior that cannot be easily reconciled with gravitational-wave scenarios (Section~\ref{sec:sou}).

Finally, we have suggested (in Section~\ref{sec:mag}) that the lack of a strong magnetosphere (able to generate a strong centrifugal barrier) in most nuclear powered pulsars 
may be related to the existence of a fast sub-population. If the magnetic field is buried, it is expected to re-emerge once accretion stops. However, field burial affects the accretion torque and one might expect that it would lead to large numbers of very fast radio ms pulsars. Since such systems are not observed, one might speculate that the magnetic field is
not only buried but also decays during the accretion phase. If this is the case, then there may exist a large unseen population of very fast spinning neutron stars.

\acknowledgments{We would like to thank R. Wijnands and C. D'Angelo for useful suggestions. AP acknowledges support from an NWO (Netherlands Organization for Scientific Research) Vidi Fellowship. }

\end{document}